\documentclass[conference]{IEEEtran}
\IEEEoverridecommandlockouts
\usepackage{cite}
\usepackage{amsmath,amssymb,amsfonts}
\usepackage{algorithmic}
\usepackage{graphicx}
\usepackage{textcomp}
\usepackage{xcolor}
\usepackage[hyphens]{url}
\usepackage[hidelinks]{hyperref}
\usepackage{subfig}
\usepackage{soul,color}
\usepackage{listings}
\newcommand{\mosaik}{\textsc{mosaik}}
\def\BibTeX{{\rm B\kern-.05em{\sc i\kern-.025em b}\kern-.08em
    T\kern-.1667em\lower.7ex\hbox{E}\kern-.125emX}}
\begin{document}

\title{Towards an Assisted Simulation Planning for Co-Simulation of Cyber-Physical Energy Systems
\thanks{The work is part of the research project 'NEDS – Nachhaltige Energieversorgung Niedersachsen', which is supported by the Lower Saxony Ministry of Science and Culture through the 'Nieders{\"a}chsisches Vorab' grant program (grant ZN3043).
\\
\copyright2019 IEEE.~https://doi.org/10.1109/MSCPES.2019.8738788
Personal use of this material is permitted. Permission from IEEE must be obtained for all other uses, in any current or future media, including reprinting/republishing this material for advertising or promotional purposes, creating new collective works, for resale or redistribution to servers or lists, or reuse of any copyrighted component of this work in other works.}
}

\author{\IEEEauthorblockN{Jan Sören Schwarz}
\IEEEauthorblockA{\textit{Department of Computing Science} \\
\textit{University of Oldenburg}\\
Oldenburg, Germany \\
jan.soeren.schwarz@uni-oldenburg.de}
\and
\IEEEauthorblockN{Cornelius Steinbrink}
\IEEEauthorblockA{\textit{R\&D Division Energy} \\
\textit{OFFIS – Institute for Information Technology}\\
Oldenburg, Germany \\
steinbrink@offis.de}
\and
\IEEEauthorblockN{Sebastian Lehnhoff}
\IEEEauthorblockA{\textit{Department of Computing Science} \\
\textit{University of Oldenburg}\\
Oldenburg, Germany \\
sebastian.lehnhoff@uni-oldenburg.de}
}

\maketitle

\begin{abstract}
Increasing complexity in the power system and the transformation towards a smart grid lead to the necessity of new tools and methods for the development and testing of new technologies.
One testing method is co-simulation, which allows coupling simulation components from different domains to test their interaction.
Because the manual configuration of complex large-scale co-simulation scenarios can be error-prone, we propose an approach for assisting the user in  the development of co-simulation scenarios.
Our approach uses an information model, a component catalog implemented in a Semantic Media Wiki, and Semantic Web technologies to assist the high-level modeling of co-simulation scenarios, recommend suitable simulation components, and validate co-simulation scenarios.
This assistance aims to improve the usability of co-simulation in the development of interdisciplinary, large-scale scenarios.
\end{abstract}

\section{Introduction}\label{introduction}
The transition towards renewable energies in the power system leads to more fluctuating and decentralized energy resources.
To handle the challenges caused by this, the traditional power system is transformed into a smart grid and new technologies have to be developed and tested.
%
Due to the integration of ICT (\textit{information and communication technology}), environmental, economic, and sociotechnical systems, testing can be quite complex.
A common approach for handling this complexity in the planning and testing is co-simulation, which allows coupling simulation models from different domains for a joint simulation \cite{Schloegl2015}.
\par
Due to the complexity of such systems, expert collaboration is of the utmost importance.
It can not be expected, however, that all experts for \textit{Cyber-Physical Energy Systems} (CPES) subsystems are co-simulation experts simultaneously.
Additionally, knowledge and terminology from different domains has to be integrated.
Therefore, tools are needed to permit the design of executable interdisciplinary setups without needing to deal directly with the details of model coupling and interfacing.
\par
The demand for an assisted or automated generation of co-simulation setups has been formulated in \cite{Steinbrink2017} and \cite{Steinbrink2018}.
It is seen as a central requirement for simulation-based testing environments in the CPES domain.
In \cite{Steinbrink2017}, this requirement is split into two desired features:
One is focused on the high-level description of co-simulation scenarios, the other deals with description standards for simulation components, which allow users to interpret them analogous to real-world components to improve coupling validation.
In this paper the focus will be on three use cases of simulation planning.
The first is the same as mentioned in \cite{Steinbrink2017}: (1) Assist domain experts in modeling high-level goals of a co-simulation scenario.
In our description of simulation planning we look at the second desired feature mentioned in \cite{Steinbrink2017} from a more active viewpoint, and split it up in (2) the assistance in the process of finding suitable simulation components, and (3) the validation of co-simulation scenarios.
To address these use cases we propose our approach using an information model for high-level planning, a catalog for co-simulation components implemented in a \textit{Semantic MediaWiki} (SMW), and Semantic Web technologies to integrate them in co-simulation. 
\par
The paper is structured as follows.
Section \ref{relWork} gives an overview of related works.
Section \ref{foundations} introduces the foundations for the approach described in section \ref{appraoch}. 
The future work is outlined in section \ref{futurework} and a conclusion given in section \ref{conclusion}.
\section{Related Work}\label{relWork}
The research project ERIGrid is focused on the improved generation and documentation of CPES test setups.
Within the project, a procedure has been developed that aims at structuring and assisting the testing process.
The so-called \textit{Holistic Testing Description} (HTD) consists of a set of templates that allow specification of testing endeavors at different stages of the process, from abstract test cases to implemented experiments \cite{VanderMeer2017}.
The HTD is currently expanded to account for the integration of statistical \textit{design of experiments} methods to ensure statistical reliability of results \cite{van2018design}.
Furthermore, a process is currently under development to generate suggestions for experiment realizations based on abstract test descriptions.
This process is based on a database for energy laboratories and their components\footnote{https://erigrid.eu/wp-content/uploads/2018/05/SchemaSpy.zip}, which can be selected based on the test requirements.
However, the process is strongly geared toward hardware experiments and displays a rather limited scope in terms of involved domains (electrical energy, communication and control).
Therefore, it cannot directly be applied to the co-simulation field that is likely to incorporate a wider range of application domains, like economic and environmental systems.
Additionally, hardware components tend to be more strongly standardized in their interfaces than simulation components, which leads to a different set of challenges in the automation of co-simulation compared to hardware experiments.
\par
In \cite{maroti2014next} a collaboration platform is presented that allows the storage of component models in a web-based database and supports domain-specific composition of complex models.
It has been shown in \cite{neema2016c2wt} how this tool can be integrated into a co-simulation platform.
However, the approach is primarily designed for the creation of \textit{Domain-Specific Languages} (DSL) and the hierarchical composition of complex models.
In CPES systems, cross-domain interactions play a bigger role than component hierarchies.
Thus, an approach is needed that aims at information flows between heterogeneous systems that otherwise act independently, following different domain constraints.
\par
The following approaches use ontologies in modeling and simulation:
Teixeira et al. \cite{Teixeira2018} 
describe the TOOCC (Tools Configuration Center) for co-simulation with integration of ontologies for the interoperability between different electricity market multi-agent simulation platforms.
But the focus of this approach seems to be mainly on energy markets and building energy management, and the ontologies seem to be mainly used as data structure of messages between simulation models and not in the development of scenarios.
Another approach is CODES (Composable Discrete-Event scalable Simulation) \cite{Teo2008}.
It contains the COSMO ontology, which supports the classification of components to allow component discovery and reuse with a model repository, but it is limited to discrete-event simulation.
Also Karhela et al. \cite{Karhela2012} describe an approach using ontologies for their modeling and simulation framework Simantics\footnote{\url{https://www.simantics.org}}.
But it is focused on industrial process simulation.
Therefore, none of these approaches can fulfill our requirements.
\begin{figure*}[!ht]
\centering
\includegraphics[width=1\textwidth]{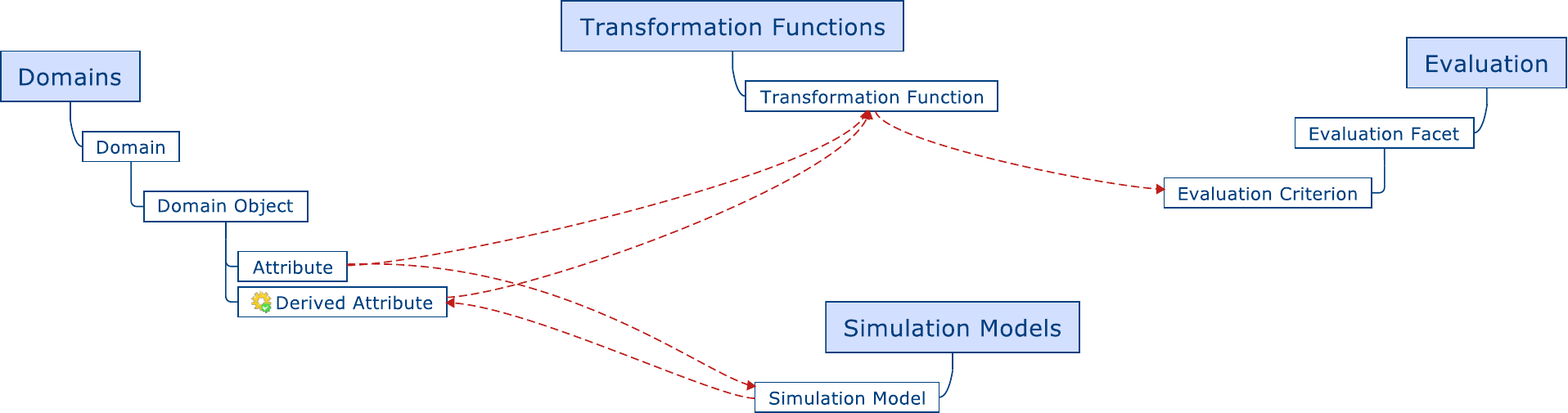}
\caption{Structure of the SEP information model \cite{CCIS2018}.}
\label{structureIM}
\end{figure*}
\section{Foundations}\label{foundations}
This section describes co-simulation in more detail, introduces Semantic Web technologies, and outlines previous work of an information model and a catalog for co-simulation components.
\subsection{Co-Simulation}
Co-simulation is defined as the coupled execution of different models, which can have different representation (e.g. ordinary/partial differential equations, discrete automata, time series), in individual runtime environments \cite{Schloegl2015}.
As described in \cite{Steinbrink2017}, co-simulation can be done based on ad-hoc coupling between the models or based on a framework coupling the models.
Because the use of a framework enables the reuse of models and this plays a crucial role for the simulation planning, we focus on frameworks.
Many different frameworks exist, which are categorized and compared in \cite{Schloegl2015} and \cite{Vogt2018}.
Our approach will be implemented as prototype with the co-simulation framework \mosaik{}\footnote{\url{https://mosaik.offis.de}} \cite{Schloegl2015,SAM2019}, which is available as open source software and is widespread in the CPES community.
It is focused on providing high usability and flexibility to enable interdisciplinary teams to develop scenarios for their tests.
\par
\mosaik{} provides two \textit{Application Programming Interfaces} (API):
The Scenario-API allows the starting and initializing of simulators, and defines the connections between different simulators to setup the scenario.
The Component-API defines the protocol for the communication between simulators and \mosaik{}.
Each simulation component is described by a so-called \textit{meta} object.
It describes primarily the models contained in the component and their parameters and attributes.
Parameters can be used to configure the model during initialization and the attributes are the interfaces for data exchange between the model and \mosaik{} during simulation.
These APIs offer a high flexibility for adding components from different domains to a simulation.
\par
The \textit{Functional Mockup Interface} (FMI) standard has been developed to allow the coupling of different simulation models in industrial and scientific projects \cite{Blochwitz2009}.
It can be seen as a more complex version of the component API, which allows more detailed description of variables.
An executable simulation model with an FMI-based description is called a \textit{Functional Mockup Unit} (FMU).
It consists of executable C code and a XML-based description of the unit.
A substantial part of the XML-based description is the definition of variables, which define the inputs and outputs of simulation models \cite{ModelicaAssociationProjectFMI2013}.
Each variable can be described by a set of attributes in FMI.
For example, the attribute causality can have values such as "input", "output", "parameter", or "calculatedParameter", or the attribute variability characterizes time instants when a variable can change its value and may have values like "constant", "fixed", "discrete", or "continuous".
\par
As described in \cite{VanderMeer2017} a coupling of \mosaik{} and FMI using the FMI++ library \cite{FMIpp} is implemented, which can be used to integrate FMUs in a \mosaik{} simulation.
In \cite{Rohjans2014} the coupling of FMI and \mosaik{} is described in more detail, but it is based on outdated versions of \mosaik{} and FMI++.
\subsection{Semantic Web}
The Semantic Web enhances the World Wide Web by adding semantic description to unstructured data \cite{Domingue2011}, but can also be used in other contexts.
By using Semantic Web technologies knowledge can be made machine readable and processable.
A common concept for the representation of domain knowledge is the use of ontologies.
An ontology describes knowledge based on concepts (classes) and their relationships (properties).
The most common format to store an ontology is the \textit{Web Ontology Language} (OWL)\footnote{\url{https://www.w3.org/TR/owl2-overview/}}.
Another essential format of the Semantic Web is the \textit{Resource Description Framework} (RDF)\footnote{\url{https://www.w3.org/RDF/}}.
It stores data in triples of subject, object, and predicate, and a database for storing such data is called a triple store.
RDF-based data can be queried with the \textit{SPARQL Protocol And RDF Query Language} (SPARQL)\footnote{\url{https://www.w3.org/TR/rdf-sparql-query/}}.
\par
The SMW~\cite{KROTZSCH2007251} is an extension for the MediaWiki\footnote{\url{https://www.mediawiki.org/wiki/MediaWiki}}, which is the software behind Wikipedia. 
With a SMW the content of a wiki can be enriched with semantic data and stored in a triple store to be accessed with SPARQL queries.
\par
In the development of ontologies it is common to integrate already existing ontologies where possible.
For the modeling of co-simulation scenarios of the power system the following ontologies are relevant.
The \textit{Ontology of units of Measure} (OM) is an OWL ontology of the domain of quantities and units of measure \cite{Rijgersberg2013}.
Thus, it can be used to model the units of attributes in the development of connections of simulation models, as units are crucial for correct data flows.
In the energy domain the \textit{Common Information Model} (CIM) is widespread to facilitate interoperability in the power system \cite{Uslar2012}.
It contains a data model in form of a domain ontology providing a vocabulary of the power system, which can be used to integrate this domain knowledge in the modeling of high-level scenarios.
\subsection{Information Model}\label{foundationsIM}
In previous work we have introduced an information model and a \textit{Sustainability Evaluation Process} (SEP) for the high-level planning of a co-simulation \cite{CCIS2018}.
The SEP describes an integrated process for the evaluation of sustainability of future scenarios based on literature review and co-simulation.
The information model supports the information exchange in the process of development and evaluation of co-simulation scenarios, as it allows modeling dependencies, data flows, and evaluation functions.
\par
The structure of the information model is depicted in \figurename \ref{structureIM} with arrows representing data flows.
On the left-hand side, the domains of interest are modeled.
Each domain can contain domain objects, which represent some real world object and are described by attributes. 
These attributes can be the input for simulation models.
The so-called derived attributes are results from a simulation model.
On the right-hand side, the evaluation function is defined.
The main function is subdivided into facets and criteria.
This evaluation function represents the results that a simulation is supposed to provide.
The connection between the two sides is established with transformation functions from attributes to the evaluation criteria. 
These functions can stand for a direct data flow, but can also represent an aggregation of data or a calculation.
\par
The content of the information model is stored as RDF to make it available for querying with SPARQL.
This RDF can be created by a tool like Prot\'{e}g\'{e}\footnote{\url{https://protege.stanford.edu/}} or by the process roughly described in \cite{EKAW2018} and \cite{CCIS2018}.
In this process the information model is modeled in a mind map to facilitate the participation of domain experts in an interdisciplinary collaboration without previous knowledge of Semantic Web technologies.
The use of a mind map initiates the development process through brainstorming and collating information step by step in the structure of the information model.
The content of the information model is transformed to RDF automatically.
To add more details than this transformation allows the RDF has to be changed directly.
\subsection{Co-Simulation Component Catalog}
\par
\begin{figure}[!ht]
\centering
\includegraphics[width=0.5\textwidth]{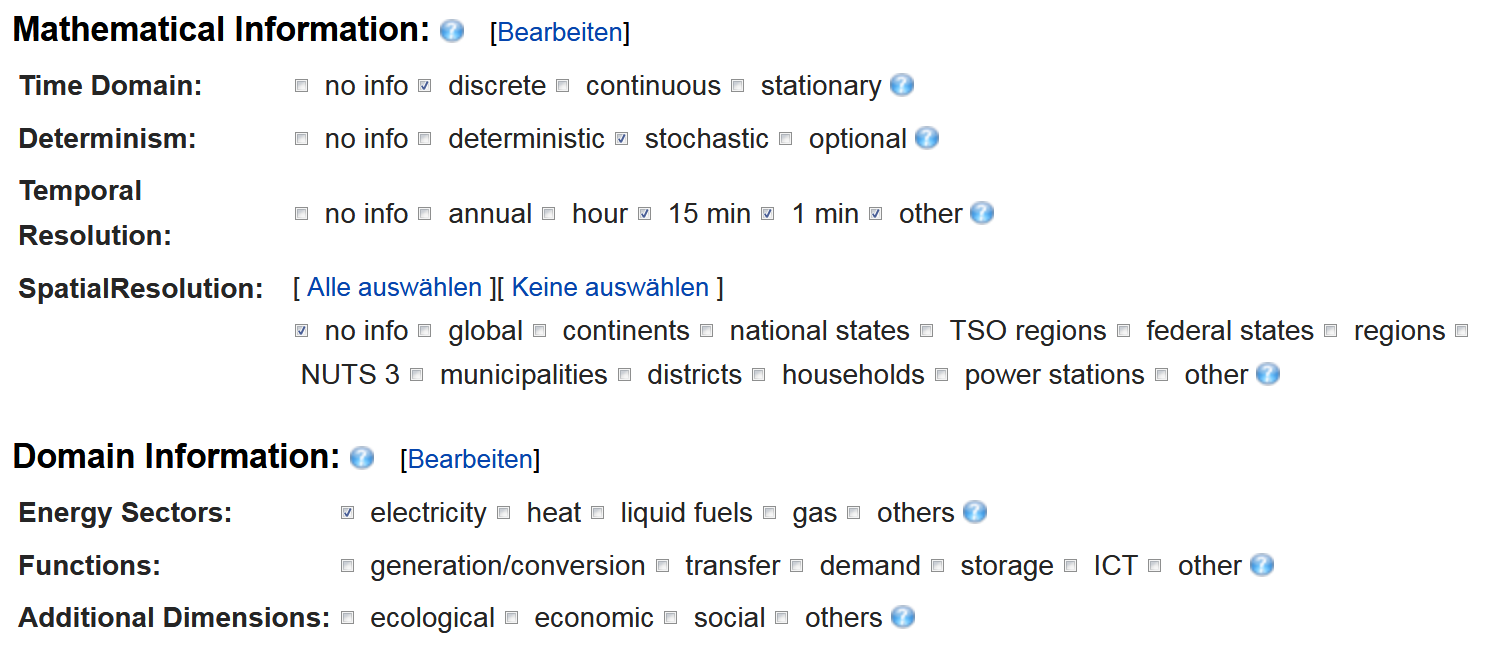}
\caption{Part of the co-simulation component catalog.}
\label{cosicoca}
\end{figure}

To specify concrete simulation scenarios applicable simulation models have to be selected and coupled based on the high-level scenarios modeled, e.g., in the information model.
To support this process a questionnaire has been developed that collects information of the available simulation components.
It has been implemented in a SMW, which was inspired by the catalog of the \textit{openmod} initiative's open models\footnote{\url{https://wiki.openmod-initiative.org/wiki/Open_Models}}. 
The catalog in the SMW is used to facilitate the participation of users.
With the \textit{Page Forms} extension\footnote{\url{https://www.mediawiki.org/wiki/Extension:Page_Forms}} it offers intuitive usable forms to edit or add components, as shown in \figurename \ref{cosicoca}.
\par
The catalog's questionnaire is subdivided into several categories:
\begin{LaTeXdescription}
	\item[General Information] describing the component, such as contact details of the developer, the type of software (E.g., simulation model, modeling framework, data analysis tool, or controller), or usage constraints.
	\item[Technical Information] describing the implementation and how it can be used and accessed.
		E.g., the hardware requirements which have to be fulfilled and which protocols and APIs the implementation provides.
	\item[Mathematical Information] describing the mathematical behavior of the component, e.g., the temporal and spatial resolution.
	\item[Domain Information] describing which domains are included in a simulation model.
		E.g., if it contains information about electricity, heat, or liquid fuels or if other domains are included (see \figurename \ref{cosicoca}). 
	\item[Variables] containing FMI definitions of variables.
\end{LaTeXdescription}
So far the catalog is only for internal usage, but a public usage is planned in the future.

\section{Simulation Planning Approach}\label{appraoch}
\begin{figure*}[hbt!]
\centering
\includegraphics[width=\textwidth]{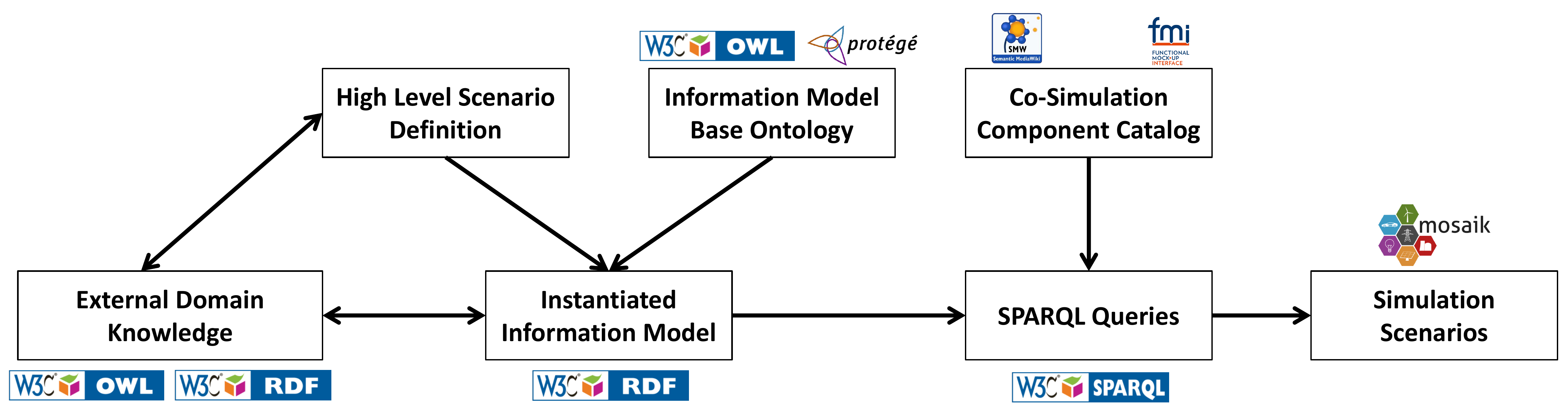}
\caption{Overview of the simulation planning process.}
\label{fig:scenarioPlanningOverview}
\end{figure*}   
Our approach for simulation planning integrates the technologies and processes introduced in the previous section  to assist the user.
The structure is based on the three use cases of simulation planning we stated in the introduction and ends with a section exhibiting the benefits of the approach for data management.
\subsection{High-level modeling of co-simulation scenarios}
The first use case is the high-level modeling of the scenario.
This means describing the goals of a simulation, identifying needed simulation models, modeling data flows, and identifying potential dependencies between simulation models.
This can be seen as preparation for the concrete implementation without going too much into the implementation details.
It also has to define the real world objects that have to be considered in the scenario.
The process of high-level modeling has to guide collaboration in an interdisciplinary project team and allows the integration of domain knowledge.
\par
We argue that co-simulation is mainly done for two different types of questions:
\par
On the one hand, co-simulation is used for strategic questions for future planning with energy scenarios \cite{Grunwald2016}.
For this strategic co-simulation the SEP with its information model described in section \ref{foundationsIM} can be used for the high-level modeling.
As shown on the left-hand side of \figurename \ref{fig:scenarioPlanningOverview} a base ontology describes the structure of the information model and makes the content available for querying after a transformation to RDF.
\par
On the other hand, co-simulation is used for operational questions, i.e. to test a new technology and its functionality.
For the high-level scenario modeling of operational simulation the information model has to be extended.
For this extension the integration with the ERIGrid HTD is planned to add definitions of hardware components, test cases, and experiments.
\subsection{Recommendation of simulation models}
Based on a high-level scenario description the next step is to develop an executable co-simulation scenario.
This means a simulation scenario is needed, which is defined by the simulation components and their connections and can be executed as simulation with \mosaik{}.
A main task for this is to find and couple suitable simulation components to fulfill the previously modeled goals.
Thus, the content of the co-simulation component catalog has to be integrated with the information model.
This way, the information from the catalog is used to support the transformation from the high-level definition of a simulation to the executable simulation scenario as shown on the right-hand side in \figurename \ref{fig:scenarioPlanningOverview}.
\par
Recommendations have to be based on the following three aspects:
\begin{LaTeXdescription}
	\item[Units:] 
		Simulation components are coupled based on their attributes.
		Thus, the units of attributes are essential for a correct coupling, because they describe the interfaces between different simulation components.
		As units can be annotated in the information model and the FMI-based definition of variables also contains units \cite{ModelicaAssociationProjectFMI2013}, this information can be used to compare the units of attributes of simulation models and the information model.
		Additionally, the OM\cite{Rijgersberg2013} can not only be used to compare the units directly, but dimensionally.
		If the units differ, but the dimension is the same, a unit transformation has to be added.
		For this, a generic unit transformation component is planned for \mosaik{}, which should be added automatically.
	\item[Topics:] 
		Two attributes with the same units do not automatically imply the same meaning as well.
		Thus, the semantic meaning of attributes also has to be considered for the coupling of simulation components. 
		Information about the semantics of the high-level modeling can be found in the domain objects in the information model or in references to external domain ontologies. 
		Information about the semantics for simulation components can be found in the catalog, e.g., the included domains and energy sectors of a components.
	\item[Dimension:] 
		For a correct execution of a co-simulation scenario the dimension of data exchanged between simulation components has to be correct.
		This means that components may have minimal or maximal values they can handle as input, which should not be exceeded. 
		In FMI this can be modeled with definition of types \cite{ModelicaAssociationProjectFMI2013}.
\end{LaTeXdescription}
\par
Recommendations for suitable simulation models can be made based on SPARQL queries accessing the information model and the catalog. 
To assist users who lack experience with SPARQL queries, predefined queries for common use cases are provided.
\subsection{Validation of simulation scenarios}
The goal of validation is to check if a concrete \mosaik{} simulation scenario is correct and coherent.
The validation has to be integrated in \mosaik{} scenario definition.
As a \mosaik{} scenario is currently implemented in a python script, the validation could be added as an additional utility function or in the connect function for simulation models.
But to achieve better flexibility an export of the scenario definition to a file is developed.
For the seamless integration with the information model and the catalog the RDF format is used.
That way, the validation can be executed with this serialized simulation scenario based on an ontology describing the \mosaik{} simulation scenario structure.
Due to the use of a standard file format like RDF, interoperability with other co-simulation tools is possible.
\par
To make the information from the catalog available in \mosaik{}, a transformation tool is planned which reads the data from the SMW and exports it in the simulation component's meta description.
\subsection{Integration in data management}
The information model can also be used for the data management of the results generated during the execution of a co-simulation scenario.
It describes the connection from the domain object attributes to the evaluation by transformation functions.
These transformation functions permit the use of multiple attributes as input and aggregate them.
So far, this is done by post-processing of the simulation results, but a utility for aggregation during simulation in \mosaik{} would improve the handling.
This aggregator has to be implemented as a \mosaik{} simulation component, which can be parametrized to aggregate its inputs.
\par
The ontology-based representation of the high-level planning allows the reuse of this information in the evaluation of the results.
Domain knowledge and terminology contained can be particularly helpful for that.
The direct integration of results in the information model can be implemented two ways:
Results can be stored, e.g., in files or a relational database, and accessed with \textit{Ontology-Based Data Management} (OBDM) tools \cite{Daraio2016}.
Results can also be stored directly in a triple store. 

\section{Future Work}\label{futurework}
The long-term goal of the proposed approach is to extend the assistance of the user to an automatic generation of concrete executable simulation scenarios from a high-level scenario definition.
To achieve this automation, the described recommendation of simulation models has to be extended to an automated coupling of simulation models.
Thus, the questionnaire of the co-simulation component catalog has to be extended with further classifiers for components, e.g., accuracy or more detailed performance and aggregation levels.
\par
To broaden the scope of high-level scenario modeling from strategic to operational and from software coupling to hardware integration, it is planned to harmonize the ERIGrid HTD procedure with the presented approach.
This will provide the scenario generation with a broader framework for documentation of high-level test cases.
As the HTD procedure is likely to be adapted and expanded in future scientific and industrial applications, this will also increase the impact of our approach. 
The advantage of the HTD is that complex test cases can be split up into manageable test specifications that can be realized by a (co-simulation) experiment each.
Furthermore, the HTD will benefit from the scenario generation through automated creation of experiments, which will make the overall process more accessible.
\section{Conclusion}\label{conclusion}
We have proposed an approach for an assisted simulation planning for co-simulation in this paper.
The description is subdivided in the following three use cases:
\par
Firstly, an information model for the high-level modeling of co-simulation scenarios is shown.
This information model has been developed for strategic simulation of future scenarios and allows modeling data flows, dependencies, and the required results of a simulation.
It is based on an ontology and its content is made available as RDF.
\par
Secondly, the integration of the information model with a co-simulation component catalog is described to recommend simulation models.
The catalog is implemented in a SMW, and describes the available components based on a questionnaire and the FMI definition of variables.
The catalog and the information model can be queried with SPARQL to assist the users in finding the best simulation models for their needs.
\par
Thirdly, the validation of executable simulation scenarios is described.
It uses questionnaire-based component description and a FMI-based variable specification from the catalog, and validates the simulation scenario based on that.
\par
Thus, the proposed approach provides the way towards a fully integrated process assisting the user in development of co-simulation scenarios and supporting interdisciplinary collaboration.

\bibliographystyle{IEEEtran}
\bibliography{IEEEabrv,bib,bibself}

\end{document}